\documentclass[twocolumn,showpacs,preprintnumbers,amsmath,amssymb]{revtex4}

\usepackage{graphicx}
\usepackage{dcolumn}
\usepackage{bm}
\usepackage[usenames]{color}

\begin{document}


\title{Ferroelectricity in multiferroic magnetite Fe$_{3}$O$_{4}$ driven 
  by noncentrosymmetric Fe$^{2+}$/Fe$^{3+}$ charge-ordering: 
  First-principles study}

\author{Kunihiko Yamauchi$^1$}
\author{Tetsuya Fukushima$^1$}
\author{Silvia Picozzi$^1$}%
 \email{silvia.picozzi@aquila.infn.it}
\affiliation{%
1. Consiglio Nazionale delle Ricerche - Istituto Nazionale di Fisica della Materia (CNR-INFM), CASTI Regional Lab., 67100 L'Aquila, Italy\\
}%

\date{\today}
\newcommand{\fo}{Fe$_{3}$O$_{4}$}
\begin{abstract}

By means of first-principles simulations, we unambiguously show that
improper ferroelectricity in magnetite in the low-temperature insulating phase is driven by
 charge-ordering.  An accurate comparison between  monoclinic ferroelectric  $Cc$ and  paraelectric $P2/c$ structures shows that 
the polarization arises because of  ``shifts" of electronic charge between octahedral Fe sites, leading to a  non-centrosymmetric  Fe$^{2+}$/Fe$^{3+}$ charge-ordered pattern.  Our predicted values for polarization, in good agreement with available experimental values, are discussed in terms of point-charge dipoles located on selected  Fe tetrahedra, pointing to a manifest example of electronic ferroelectricity driven by charge rearrangement.  

\end{abstract}

\pacs{Valid PACS appear here}
\maketitle


``Improper multiferroics''\cite{maxim,kimura,efremov,prlslv,iGK} are attractive multifunctional materials, where magnetism and ferroelectricity are strongly coupled.
They can be classified according to different  driving forces, which primarily break the spatial inversion symmetry paving the way to ferroelectricity : 1) spin order, 2) charge order (CO) and 3) orbital order. In this letter, we will focus on the second group, where LuFe$_2$O$_{4}$ has emerged as a prototype.\cite{ikeda} However, recently
 Fe$_{3}$O$_{4}$ has been also suggested as a ferrimagnet with ferroelectricity being induced by charge-ordering,\cite{Khomskii} which would depict magnetite as the first multiferroic known to mankind.
\fo  shows the well-known first order metal-insulator (Verwey) transition at T$_{V}\sim$120K, below which the crystal structure changes from cubic $Fd3m$ to monoclinic structure and Fe$^{2+}$/Fe$^{3+}$ charge ordering is observed at the Fe B sites in the inverse-spinel AB$_{2}$O$_{4}$ lattice.\cite{wright,schlappa} 
Earlier experiments\cite{kato} suggested Fe$_{3}$O$_{4}$ to show a spontaneous polarization 
at low temperatures, with the structure  possibly undergoing a transition from  monoclinic to  triclinic symmetry\cite{medrano}. However, up to date the physical mechanism underlying the rising of ferroelectricity is largely unknown.
Khomskii\cite{Khomskii} has suggested that ferroelectric (FE) polarization is caused by a combination of site-centered and bond-centered charges between Fe$^{2+}$/Fe$^{3+}$ ions in the CO B-site Fe$_4$-tetrahedron pattern. His model, based on the structure of Ref.\cite{wright}  assumes each tetrahedron to show a ``3:1'' CO arrangement (three  Fe$^{2+}$ and one Fe$^{3+}$ ions in a tetrahedron, or viceversa), at variance with Anderson's criterion,\cite{Anderson} where each tetrahedron shows a ``2:2'' pattern (two  Fe$^{2+}$ and two Fe$^{3+}$ ions). 
The 2:2 arrangement was found from  density functional theory (DFT)\cite{Piekarz} on the cubic $Fd3m$ structure, distorted by $X_3$ phonon mode. 
Moreover, the 3:1 arrangement was recently obtained within DFT\cite{Piekarz,leonov,Jeng_prl} on a monoclinic $P2/c$ structure. In addition, a mixed CO pattern (25\% of 2:2 and 75\% of 3:1 tetrahedra) was suggested to occur in a base-centered monoclinic $Cc$ structure\cite{Jeng_prb}.
Our DFT results, in agreement with Ref.\cite{Jeng_prb},
predict the $Cc$ structure to be the ground state, consistently with recent results from resonant X-ray scattering\cite{rixs}. As discussed in Ref.\cite{Jeng_prb,wright_prl}, the  $Cc$ symmetry is stabilized by a delicate balance of different effects: Coulomb repulsion, entropy,\cite{Anderson}, Fermi-surface nesting \cite{yanase} leading to a [001] charge-density wave (CDW),\cite{wright_prl} oxygen breathing modes, etc.

Recent experimental data showing real time FE switching in magnetite epitaxial thin films were  found to be in good agreement with our DFT calculations for the polarization {\bf P} in  the $Cc$ FE structure.\cite{Alexe} 
In this letter, we focus on the proof of charge ordering as the microscopic origin of ferroelectricity in \fo, by comparing two monoclinic $P2/c$ and $Cc$ lattice structures.  
Since the  $P2/c$ structure shows inversion symmetry, the total {\bf P}  must cancel out, so that at most antiferroelectricity may occur; on the other hand, the $Cc$ structure is ferroelectrically active due to the lack of centrosymmetry. 

Electronic structure calculations were performed using the  ``Vienna $Ab$ $initio$ Simulation Package (VASP)'',\cite{vasp} 
where the projector-augmented-wave potentials, the  generalized gradient approximation  to the exchange-correlation potential\cite{pbe} plus an effective on-site Coulomb interaction $U$ were used.\cite{ldau}
In order to compare the total energies, the FE polarization and other relevant properties,  we used - for both paraelectric (PE) $P2/c$ and FE  $Cc$ states - the primitive cell of the base-centered monoclinic $Cc$ lattice 
(with 112 atoms/cell) and experimental lattice parameters\cite{Zuo}. Since the difference in experimental lattice constants between the PE and FE phases is $\sim$0.1\%, the same lattice parameters in both  unit-cells were used.
The polarization vector {\bf P} is described with the conventional lattice vectors ($a,b,c$).
Internal atomic coordinates  were optimized in both $P2/c$ and $Cc$,  starting from experimental Wyckoff parameters\cite{wright, Zuo}. The conventional atomic positions of the $Cc$ lattice were displaced by (-1/8, 0, 0) to fit into the $P2/c$ structure, so as to have the pseudo-inversion center at (0, 0, 0).  We focus on the $P2/c$ and $Cc$ structures since they are both 
proposed in diffraction experiments\cite{wright, Zuo} and show the lowest DFT total energies compared to other proposed symmetries, {\em i.e.} $Pmca$ and $Pmc2_1$.\cite{Jeng_prb} Note that the base-centered monoclinic lattice of $Cc$ ($a=b\neq c$, $\alpha \neq \beta \neq \gamma \neq 90^{\circ})$ is almost identical to a triclinic lattice (experimentally suggested  as the FE state),  the only difference being the $a=b$ condition.
A cutoff energy of 400 eV for plane waves, $4\times4\times2$ Monkhorst-Pack $k$-point grid in the Brillouin zone, a threshold of atomic forces of  0.03 eV/\AA$\:$ were used.  An effective Coulomb energy $U=4.5$ eV and an exchange parameter $J=0.89$ eV (as in Ref.\cite{Jeng_prl}) were used for Fe-$d$ states, although some  calculations with different values of $U$ were performed (see below).
Fe-$3p^{6}3d^{6}4s^{2}$ and O-$2s^{2}2p^{4}$ electrons were treated as valence states. 
For Fe ions, the ferrimagnetic configuration was considered, with all octahedral Fe sites as up-spin sites and all tetrahedral Fe sites as down-spin sites. 
Spin-orbit coupling was neglected.
%

In terms of relevant structural and electronic properties, our results are similar to the previous study by Jeng et. al. \cite{Jeng_prb}, 
so in this letter we will focus only on FE properties.  
\begin{figure}
\resizebox{82mm}{!}
{
\includegraphics{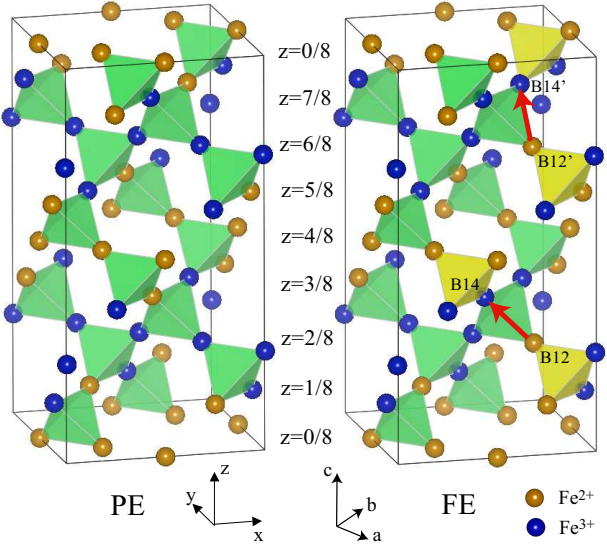}
}
\caption{\label{fig:lattice} 
Ionic structure of Fe octahedral sites in $P2/c$ (left) and $Cc$ (right) cells. Orange and blue balls show Fe$^{2+}$ and Fe$^{3+}$ ions, respectively. Fe$_{4}$ tetrahedra of 2:2 and 3:1 CO patterns are highlighted by yellow and green color planes, respectively. Electric dipole moments caused by charge shifts are indicated by red arrows. The  lattice vectors in the primitive unit cell ($x$,$y$,$z$) and the conventional cell ($a$,$b$,$c$) are indicated.}
\end{figure}
\begin{figure}
\resizebox{82mm}{!}
{
\includegraphics{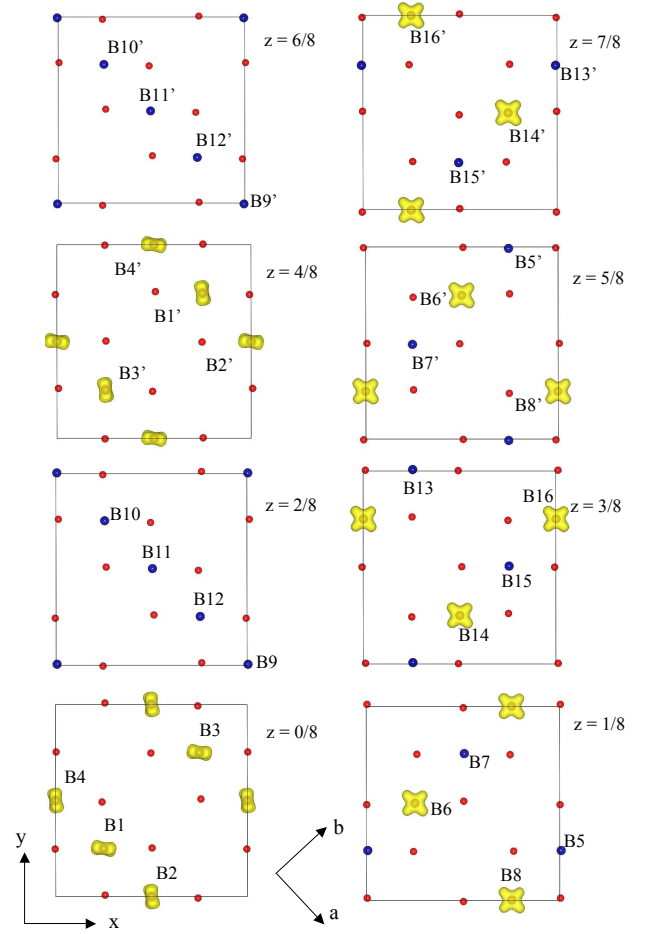}
}
\caption{\label{fig:secchg_PE} 
Charge/orbital ordering of Fe minority $t_{2g}$ states in PE configuration. Blue (orange) large balls show Fe$^{3+}$ (Fe$^{2+}$) ions. Red small balls show O ions.}
\end{figure}
\begin{figure}
\resizebox{82mm}{!}
{
\includegraphics{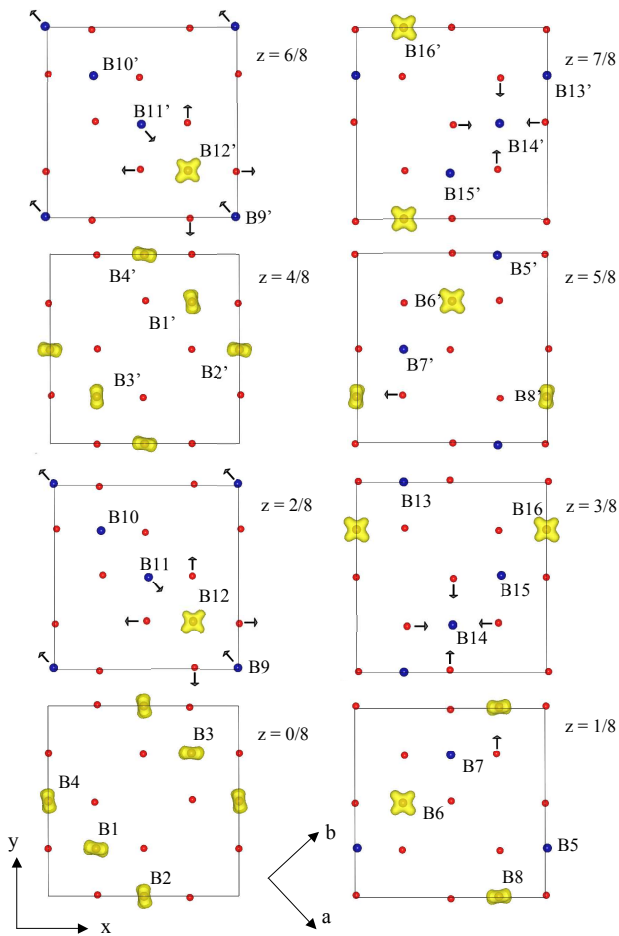}
}
\caption{\label{fig:secchg_FE} 
Charge/orbital ordering of Fe minority $t_{2g}$ states in FE configuration (notation as in Fig.\ref{fig:secchg_PE}). Black arrows show atomic displacements with respect to the PE reference state.
 }
\end{figure}
As shown in Fig.\ref{fig:lattice}, octahedral Fe sites are located in $xy$ planes with $z$ = i/8 (i=0...7). 
The PE state has \{$E, C_{2b}+(0,0,1/2), I, \sigma_{2b}+(0, 0, 1/2) $\} symmetries and the
FE state has \{$E, \sigma_{2b}+(0, 0, 1/2) $\} symmetries in a conventional base centered monoclinic cell so that there are two equivalent atoms (cfr. B12 and B12' sites in Fig.\ref{fig:lattice}). 
Note that {\em i}) the mirror symmetry along with the translation vector forbids any net polarization along $b$ and finite {\bf P} is allowed only along the $a$ and $c$ directions; {\em ii}) the translation vector is relevant for the CDW stabilization and for opening the energy gap.\cite{wright_prl} 
As discussed in Ref.\cite{Jeng_prl, Jeng_prb}, the PE state shows entirely a 3:1 tetrahedron CO arrangement whereas the
FE state shows a mixed pattern. 
The difference between the two CO distributions   (see Fig. \ref{fig:secchg_PE} for the $P2/c$ and Fig. \ref{fig:secchg_FE} for the $Cc$) can be understood when assuming a charge ``shift" from B12 to B14 site
and, in the upper part of the cell, from B12' to B14', all the other sites keeping their valence state.
Each charge shift creates two 2:2 CO tetrahedra, so as to form four 2:2 tetrahedra  (cfr. Fig.\ref{fig:lattice}). The resulting CO pattern
lacks inversion symmetry, therefore  allowing FE polarization. 
The CO rearrangement  implies a change in the local breathing mode of O ions, which are driven away (attracted) by the substituted Fe$^{2+}$ ( Fe$^{3+}$) ion at B12 (B14) site, with a displacement of $\sim0.1$\AA (cfr small black arrows in Fig. \ref{fig:secchg_FE}).
Similarly, upon charge shifts of the Fe $t_{2g}$ electron between sites, some of  the Fe$^{3+}$ ions (at B9 and B11 sites) move towards the Fe$^{2+}$ ions, maybe reminiscent of 
Khomskii's mixed bond-/site- centered charge mechanism.\cite{Khomskii} 
However, as far as ferroelectricity is concerned, when focusing on a Fe chain along [100], these movements average out so as not to give a net  contribution to {\bf P}.

As for calculated {\bf P} (cfr Table \ref{tbl.Psummary}),  the Berry phase approach\cite{berry}
 predicts  quite a large polarization (at least compared to other improper multiferroics\cite{maxim,kimura}), its
 direction lying in the $ac$ mirror plane. 
 As noted previously\cite{Alexe}, the DFT results are in excellent agreement with recently reported experimental values for magnetite thin films
 ($P\sim$ 5.5 $\mu C/cm^2$ in the $ab$ plane with the $c$ component not measured) as well as with 
  earlier experiments on single crystals\cite{kato}: $P_a$ = 4.8 $\mu C/cm^2$ and $P_c$ = 1.5 $\mu C/cm^2$. 

\begin{table}[h] \begin{center}
\caption{{\bf P}  along $a$, $b$ and $c$ axes (in $\mu$C/cm$^{2}$), calculated with different approaches (see text). } 
\label{tbl.Psummary}
\begin{tabular}{ccc}
\hline 
{\bf P$_{Berry}$} & {\bf P$_{PCM}$} & {\bf P$_{dip}$}\\
      (-4.41, 0, 4.12)& ( -4.20, 0, 5.27) & (-4.05, 0, 5.73)\\
\hline 
\end{tabular}
\end{center}
\end{table}

To deepen our analysis, we compare the value of {\bf P$_{Berry}$}   with the value estimated from a point charge model, ({\bf P$_{PCM}$}), assuming the nominal valence for every ion, {\em i.e.} ``full" charge disproportionation, {\em i.e.} 2+ and 3+ for charge-ordered Fe, 2- for O)
as well as with a simple model ({\bf P$_{dip}$})
  where we summed up the two dipole moments located at  sites where charge-shifting occurs ({\em i.e.} from B12 to B14 site) with nominally one electron (see red arrows in Fig. \ref{fig:lattice}). 
The  consistency of these values, shown in Table \ref{tbl.Psummary}, implies two important facts: ({\em i}) {\bf P}
 is induced largely by the CO rearrangement ({\em i.e.} charge shifts at few sites) but not much by the ionic displacements,  as evident by the similarity between {\bf P$_{PCM}$}  and {\bf P$_{dip}$}; ({\em ii}) since the Berry phase value is equivalent to the sum of
Wannier-function (WF) centers\cite{berry}, the similarity between {\bf P$_{Berry}$} and {\bf P$_{PCM}$}   suggests the WF of Fe-$d$ electron to be mostly centered at the ionic sites, no matter how the WF is delocalized and hybridized with surrounding O-$p$ orbitals.

\begin{table}[h] \begin{center}
\caption{Charge separation (cs, {\em i.e.} difference of $d$-charges between Fe$^{2+}$ and Fe$^{3+}$ ions in the atomic sphere with 1\AA $\:$ radius) and the corresponding FE {\bf P$_{Berry}$}  (in $\mu$C/cm$^{2}$) vs  Coulomb repulsion $U$ ($J$ is fixed to 0.89 eV).}
\label{tbl.Udependence}
\begin{tabular}{cccc}
\hline
$U$ (eV)& 4.5 & 6.0 & 8.0 \\
\hline
cs&           0.17&0.23&0.30\\
{\bf P$_{Berry}$} &       (-4.41, 0, 4.12)&(-4.42, 0, 4.81)&(-4.33, 0, 5.07)\\
\hline
\end{tabular}
\end{center}
\end{table}
Our picture based on the WF centers is also confirmed by the results of {\bf P$_{Berry}$} values upon varying the $U$ Coulomb parameter. 
As shown in table \ref{tbl.Udependence}, upon increasing $U$ and keeping the  atomic configuration fixed to that obtained for $U$ = 4.5 eV,
the charge separation between Fe$^{2+}$ and Fe$^{3+}$ is enhanced. Note that the value of {\bf P} does not change rapidly with $U$,  suggesting
the centers of the WF not to move significantly far from the  Fe$^{2+}$  for all $U$ values. Also, in the limit
of extemely large $U$, we expect a ``full" charge disproportionation to occur, causing {\bf P$_{Berry}$} to become progressively closer to {\bf P$_{PCM}$} (as confirmed by Table \ref{tbl.Udependence}).

\begin{figure}
\resizebox{82mm}{!}
{
\includegraphics{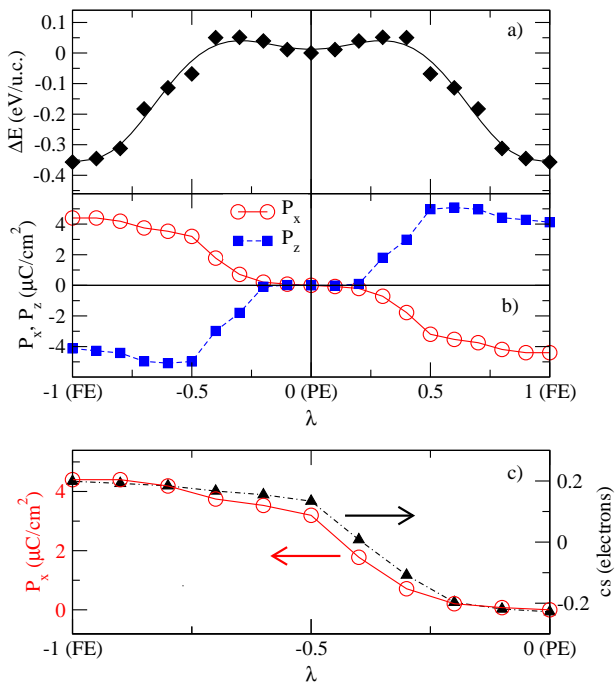}
}
\caption{a) Total energy per unit cell ($\Delta$E, with the black line as a guide to the eye) and b) Polarization ($P_x$ and $P_z$ components)  along the adiabatic path between FE structures with opposite {\bf P} through the PE structure, vs the scaling parameter $\lambda$ (see text). c) Polarization $P_x$ (left scale on the $y$-axis, see empty red circles) and charge separation (right scale on the $y$-axis and filled black triangles) along the FE ($\lambda$ =-1) to PE ($\lambda$=0) path.}
\label{fig:path} 
\end{figure}

For further insights,
an ``adiabatic" path is set up to connect the PE and FE state by displacing all ions linearly with a scaling parameter $\lambda$ ({\em i.e.} $\lambda$=1 for full FE displacements and $\lambda$=0 for the initial PE structure). The FE structure with opposite {\bf P} 
was built starting from the PE structure and with  $\lambda$=-1, {\em i.e.} with displacements opposite to the structure considered so far.  
By analogy with the above discussion for positive $\lambda$ values, two charge shifts between B10  and B6 (B10' and B6') occur as a transition from the PE to the ``negative" FE phase. 
The charge separation (cs) is calculated by comparing the ``muffin-tin" charge on the sites where the shift occurs in going from PE to FE (i.e. B12 - B14 sites when $\lambda >$ 0). The results are reported in Fig.\ref{fig:path}. The total energy shows a double valley structure, typical for ferroelectricity, with a deep global minimum at the FE state, along with 
a very shallow local minimum at the PE state.
In going from the PE to the FE state, we show that the charge shift rather suddenly occurs around $\lambda$=0.5, when  {\bf P} is strongly enhanced and the total energy reduced (cfr. Fig.\ref{fig:path} a) and b)). Moreover, the origin of ferroelectricity as driven by charge rearrangements on different sites is clearly confirmed by the trend of $P_x$ closely following that of cs along the path ((cfr. Fig.\ref{fig:path} c)). 

In summary,  we have shed light on the microscopic origin of ferroelectric polarization in insulating magnetite by analyzing the differences between two charge-ordered states: the $P2/c$ PE state and the $Cc$  FE state. Ferroelectricity is induced by a non-centrosymmetric charge-ordering and the polarization is primarily caused by local dipoles 
at selected octahedral sites, pointing to a picture of ferroelectricity mostly based on localized ``charge shifts". Our calculations  show that \fo $\:$ can be considered as a prototypical case of charge-order driving multiferroicity
with relatively large values for the electric polarization.

\acknowledgments
We thank Prof. Daniel Khomskii for  his careful reading of the manuscript.
The research leading to these results has received funding from the European Research Council under the EU Seventh Framework Programme  (FP7/2007-2013) / ERC grant agreement n. 203523.
Computational support from Caspur  Supercomputing Center  (Rome) is gratefully acknowledged. 



\end{document}